\documentclass[12pt]{article}

\usepackage{graphicx}

\def\beq   {\begin{equation}}
\def\eeq   {\end{equation}}
\def\bea {\begin{eqnarray}}
\def\eea {\end{eqnarray}}

\def\psla{p\kern-.45em/}

\def\sq{\ifmmode{\tilde{q}} \else{$\tilde{q}$} \fi}
\def\sg{\ifmmode{\tilde{g}} \else{$\tilde{g}$} \fi}
\def\tchi{\ifmmode{\tilde{\chi}} \else{$\tilde{\chi}$} \fi}

\def\ms{\ifmmode{\overline{\rm MS}} \else{$\overline{\rm MS}$} \fi}

\begin{document}


\vspace*{-1cm} 
\begin{flushright}
  TU-796 
\end{flushright}

\vspace*{1.4cm}

\begin{center}

{\Large {\bf
$b\to s\nu\bar{\nu}$ decay in the MSSM: 
Implication of $b\to s\gamma$ at large $\tan\beta$ 
} 
}\\

\vspace{10mm}

{\large Youichi~Yamada}

\vspace{6mm}
\begin{tabular}{l}
{\it Department of Physics, Tohoku University, Sendai 980-8578, Japan}
\end{tabular}

\end{center}

\vspace{1.5cm}

\begin{abstract}
\baselineskip=15pt
The decay $b\to s\nu\bar{\nu}$ is discussed in the minimal supersymmetric 
standard model with general flavor mixing for squarks, at large $\tan\beta$. 
In this case, in addition to the chargino loop contributions 
which were analyzed in previous studies, 
$\tan\beta$-enhanced contributions from the 
gluino and charged Higgs boson loops might become sizable compared with 
the standard model contribution, at least in principle. 
However, it is demonstrated that the experimental bounds on the 
new physics contributions to the radiative decay $b\to s\gamma$ should 
strongly constrain these contributions to $b\to s\nu\bar{\nu}$, 
especially on the gluino contribution. 
We also briefly comment on a possible constraint from 
the $B_s\to\mu^+\mu^-$ decay. 
\end{abstract}

\vspace{20mm}

\newpage
\pagestyle{plain}
\baselineskip=15pt

\section{Introduction}
\label{sect-intro}
%
Recently, there has been significant experimental improvements in 
the measurements of flavor-changing neutral current (FCNC) processes 
of $B$ mesons at B factories and Tevatron. 
For the $b\to s$ transition, 
experimental data for $b\to s\gamma$ and $b\to sl^+l^-$ ($l=e,\mu$) decays, 
$B_s-\bar{B}_s$ oscillation, and 
$B_s\to\mu^+\mu^-$ decay have already started to constrain 
possible contributions from new physics beyond the standard model. 

Here we focus our attention to one of the $b\to s$ processes, 
the decay into neutrino pairs \cite{bsnunusm,bsnunusm2}, 
\beq
b\to s\nu\bar{\nu}.    \label{eq-bsnunu}
\eeq 
It is known that the decays of the B mesons induced by the partonic 
process (\ref{eq-bsnunu}), especially the inclusive 
branching ratio ${\rm BR}(\bar{B}\rightarrow X_s \nu\bar{\nu})$, 
have small theoretical uncertainty 
due to the absence of photonic penguin and 
strong suppression of light quark contributions. 
On the other hand, experimental search of the 
decay (\ref{eq-bsnunu}) is a hard task. At present, only the upper 
bounds are known for both inclusive \cite{bsnunuexp1} and 
exclusive \cite{bsnunuexp4} branching ratios, at 90\% C.L., 
\bea
&& \sum_{\nu}{\rm Br}(\bar{B}\to X_s\nu\bar{\nu})<6.4\times 10^{-4}, 
\nonumber \\
&& \sum_{\nu}{\rm Br}(B^+\to K^+ \nu\bar{\nu})<1.4\times 10^{-5},\nonumber \\
&& \sum_{\nu}{\rm Br}(\bar{B}^0\to K^0_S \nu\bar{\nu})<1.6\times 10^{-4},
\nonumber \\
&& \sum_{\nu}{\rm Br}(\bar{B}^0\to K^{*0}\nu\bar{\nu})<3.4\times 10^{-4},
\nonumber\\
&& \sum_{\nu}{\rm Br}(B^+\to K^{*+} \nu\bar{\nu})<1.4\times 10^{-4}, 
\eea
which are still one order of magnitude larger than the 
standard model predictions for the inclusive \cite{bsnunusm3} and 
exclusive \cite{bsnunusm4} modes, 
\bea 
&& \sum_{\nu}{\rm Br}(\bar{B}\to X_s\nu\bar{\nu})_{SM}= 
(3.7\pm 0.2)\times 10^{-5}, \nonumber \\
&& \sum_{\nu}{\rm Br}(\bar{B}\to K\nu\bar{\nu})_{SM}= 
(3.8^{+1.2}_{-0.6} )\times 10^{-6}, \nonumber \\
&& \sum_{\nu}{\rm Br}(\bar{B}\to K^*\nu\bar{\nu})_{SM}= 
(1.3^{+0.4}_{-0.3} )\times 10^{-5}.
\label{eq-bsnunusm} 
\eea 
Future upgrades of the B factories \cite{superB} will 
extend the search region for the exclusive decays. 
For example, ${\rm Br}(B^+\to K^+ \nu\bar{\nu})$ around the level of 
the standard model prediction (\ref{eq-bsnunusm}) 
is expected to be measured at the precision of 20\% with integrated 
luminosity 50--100 ab$^{-1}$. 
On the other hand, a future $e^+e^-$ collider running on the $Z$-boson 
resonance (GIGA-Z) has a potential \cite{gigaz} 
to produce very large number of $Z\to b\bar{b}$ events, and 
possibility to greatly improve previous studies of 
the inclusive modes \cite{bsnunuexp1} at the LEP I, 
to measure the inclusive branching ratio. 
%
%

In this paper, we consider the decay (\ref{eq-bsnunu}) 
in the framework of the minimal supersymmetric 
standard model (MSSM) \cite{mssm} with general flavor mixing of squarks, 
and study the contributions of new particles, namely the 
supersymmetric (SUSY) particles and Higgs bosons. 
In cases where the value of $\tan\beta$, the ratio of the vacuum 
expectation values of two Higgs boson doublets in the MSSM, 
is not much larger than unity, 
it is shown \cite{BBMR,bsnunuSUSY} that the chargino-squark loops 
give main part of the new physics contributions to the decay, 
and may become sizable when large flavor mixing 
is present in the left-right mixing part of the up-type squark mass 
matrix. Note that this is also the case for the 
SUSY contributions to the related 
decays $K\to\pi\nu\bar{\nu}$ \cite{Kpinunu0,Kpinunu}. 

At large $\tan\beta$, say similar to or larger than $m_t/m_b\sim 40$, 
the MSSM loop contributions other than charginos might become 
also important, at least in principle. For example, gluino-squark loop 
contributions are generated by $\tan\beta$-enhanced large left-right 
mixing of down-type squarks. When, in addition, sizable mixing 
between down-type squarks in the second and third generations are present, 
gluino contribution might become sizable. 
It is also possible that, as explained later, charged Higgs boson
might give sizable loop contributions due to the 
flavor-changing effective Higgs-quark couplings, 
generated by $O(\tan\beta)$ SUSY loop corrections, 
as pointed out in Ref. \cite{Kpinunu2} for the $K\to\pi\nu\bar{\nu}$ decays. 
However, parameters in the SUSY and Higgs sectors should receive 
stringent constraints from existing measurements of the FCNC processes, 
which might suppress possible magnitudes of their contributions to 
$b\to s\nu\bar{\nu}$. 
In this paper, we will present a rough estimate of the possible 
constraints from the decay $b\to s\gamma$, by showing 
correlations between the new physics contributions 
to the Wilson coefficients for $b\to s\nu\bar{\nu}$ and those 
for $b\to s\gamma$, for each SUSY/Higgs sector separately. 
We will also comment on the implication of the $B_s\to\mu^+\mu^-$ decay 
to the Higgs boson contributions. 

The paper is organized as follows. 
In Sec.~\ref{sect-bsnunu}, we present basic formulas for 
the analysis of the $b\to s\nu\bar{\nu}$ decay in the MSSM. 
In Sec.~\ref{sect-MSSMcontr}, numerical results for the 
new physics contributions in the MSSM to $b\to s\nu\bar{\nu}$ 
are presented as correlations with those to $b\to s\gamma$ for 
each new physics sector. 
An additional constraint from the 
$B_s\to\mu^+\mu^-$ decay on the Higgs boson 
contributions is briefly commented in Sec.~\ref{sect-bsmumu}. 
Finally, conclusion is given in Sec.~\ref{sect-conclu}.

\section{$b\to s\nu\bar{\nu}$ decay in the MSSM}
\label{sect-bsnunu}

The $b\to s\nu\bar{\nu}$ decay is described by the effective Hamiltonian, 
in the notation of Ref.~\cite{bsnunusm4}, 
\beq 
H_{\rm eff} = -\frac{4G_F}{\sqrt{2}}
K^*_{ts}K_{tb} 
[C_{\nu}{\cal O}_L + C'_{\nu}{\cal O}_R ], 
\label{Heff}
\eeq
where $K_{ij}$ is the Cabibbo-Kobayashi-Maskawa (CKM) 
matrix. Here the relevant operators are 
\bea 
{\cal O}_L &=& 
\frac{\alpha}{2\pi}(\bar{s}_L\gamma^{\mu}b_L)(\bar{\nu}_L\gamma_{\mu}\nu_L), 
\label{eq-onunuL}
\\
{\cal O}_R &=& 
\frac{\alpha}{2\pi}(\bar{s}_R\gamma^{\mu}b_R)(\bar{\nu}_L\gamma_{\mu}\nu_L) .
\label{eq-onunuR}
\eea
The inclusive branching ratio is then expressed in terms of the 
Wilson coefficients ($C_{\nu}$, $C'_{\nu}$) in Eq.~(\ref{Heff}) 
as \cite{bsnunuSUSY}
\beq
\sum_{\nu}{\rm Br}(\bar{B}\to X_s\nu\bar{\nu})\sim  
\frac{N_{\nu}\alpha^2}{4\pi^2}
{\rm Br}(\bar{B}\to X_c e\bar{\nu}_e)\frac{|K_{tb}K^*_{ts}|^2}{|K_{cb}|^2}
(|C_{\nu}|^2 + |C'_{\nu}|^2 ) ,
\label{eq-inclwidth}
\eeq 
up to the QCD corrections and 
$O(m_c^2/m_b^2)$ corrections to the semileptonic decays 
$\bar{B}\to X_c e\bar{\nu}_e$. Interference between $C_{\nu}$ and 
$C'_{\nu}$ appears in the branching ratios of the exclusive modes 
$\bar{B}\to(K\nu\bar{\nu},K^*\nu\bar{\nu},\cdots)$ \cite{bsnunusm2,bsnunusm4}. 
Note that, in the massless quark limit, $(C_{\nu},C'_{\nu})$ are 
independent of the renormalization scale in QCD. 

In the MSSM, the interaction (\ref{Heff}) 
is generated by the $Z$-boson penguin and box diagrams. 
The standard model particles only contribute to 
$C_{\nu}$, giving at the leading order 
in QCD \cite{inamilim,BBMR,bsnunusm,bsnunusm2,bsnunusm4}, 
\beq
C_{\nu,\rm SM} = -\frac{1}{\sin^2\theta_W} \frac{x}{8(x-1)^2}
\left[ x^2+x-2+3(x-2)\log x \right] , 
\eeq
where $x=m_t^2/m_W^2$. 
Numerically, $C_{\nu,\rm SM}$ is about $-6.8$ for $m_t=171$ GeV. 

New particles in the MSSM, namely the SUSY particles and Higgs bosons, 
may contribute to both $C_{\nu}$ and $C'_{\nu}$, 
\bea
&&
C_{\nu}=C_{\nu,{\rm SM}}+C_{\nu}({\rm new}), \;\;\; 
C'_{\nu}=C'_{\nu}({\rm new}) , \nonumber \\
&& 
C^{(')}_{\nu}({\rm new}) = C^{(')}_{\nu,\sg} 
+ C^{(')}_{\nu,\tchi^{\pm}}+C^{(')}_{\nu,\tchi^0}+C^{(')}_{\nu,H^{\pm}}. 
\eea
$C^{(')}_{\nu}({\rm new})$ consists of the contributions 
of the gluino $\sg$ - down type squark loops, chargino 
$\tchi^{\pm}$ - up-type squark loops, neutralino $\tchi^0$ - down-type squark 
loops, and charged Higgs boson $H^{\pm}$ - top quark loops. 
Below we list the analytic forms of these one-loop contributions 
for each sector: 
\bea
C_{\nu,\sg} &=& 
-\frac{4g_s^2}{3e^2K^*_{ts}K_{tb}}
(\Gamma^{\dagger}_{DL})_{2i}(\Gamma_{DR})_{ik}
(\Gamma^{\dagger}_{DR})_{kj}(\Gamma_{DL})_{j3}
C_{24}(\tilde{d}_i,\tilde{d}_j,\sg) , 
\label{clnugl}
\\
C'_{\nu,\sg} &=& 
\frac{4g_s^2}{3e^2K^*_{ts}K_{tb}}
(\Gamma^{\dagger}_{DR})_{2i}(\Gamma_{DL})_{ik}
(\Gamma^{\dagger}_{DL})_{kj}(\Gamma_{DR})_{j3}
C_{24}(\tilde{d}_i,\tilde{d}_j,\sg) , 
\label{crnugl}
\\
C_{\nu,\tchi^{\pm}} &=& 
\frac{a^{C*}_{ik2}a^C_{jl3}}{2e^2K^*_{ts}K_{tb} } 
\left[ -\delta_{kl}(\Gamma_{UL})_{i\gamma}
(\Gamma_{UL}^{\dagger})_{\gamma j} C_{24}(\tilde{u}_i,\tilde{u}_j,
\tilde{\chi}^{\pm}_k) \right. \nonumber\\ 
&& \left. 
+\delta_{ij}V^*_{k1}V_{l1}
\{ C_{24}(\tilde{u}_i,\tilde{\chi}^{\pm}_k,\tilde{\chi}^{\pm}_l) 
-\frac{1}{4} \} 
-\frac{1}{2}\delta_{ij}U_{k1}U^*_{l1}
m_{\tilde{\chi}^{\pm}_k}m_{\tilde{\chi}^{\pm}_l}
C_0(\tilde{u}_i,\tilde{\chi}^{\pm}_k,\tilde{\chi}^{\pm}_l) \right] \nonumber\\ 
&& 
+ \frac{a^{C*}_{ik2}a^C_{il3}m_W^2}{2e^2K^*_{ts}K_{tb} } 
U_{k1}U^*_{l1}m_{\tilde{\chi}^{\pm}_k}m_{\tilde{\chi}^{\pm}_l}
D_0(\tilde{u}_i,\tilde{\chi}^{\pm}_k,\tilde{\chi}^{\pm}_l,\tilde{l}^-) , 
\label{clnuch}
\\
C'_{\nu,\tchi^{\pm}} &=& 
\frac{b^{C*}_{ik2}b^C_{jl3}}{2e^2K^*_{ts}K_{tb} } 
\left[ \delta_{kl}(\Gamma_{UR})_{i\gamma}
(\Gamma_{UR}^{\dagger})_{\gamma j} C_{24}(\tilde{u}_i,\tilde{u}_j,
\tilde{\chi}^{\pm}_k) \right. \nonumber\\ 
&& \left. 
+\delta_{ij}U_{k1}U^*_{l1}\{ C_{24}(\tilde{u}_i,\tilde{\chi}^{\pm}_k,
\tilde{\chi}^{\pm}_l) -\frac{1}{4} \} 
-\frac{1}{2}\delta_{ij}V^*_{k1}V_{l1}
m_{\tilde{\chi}^{\pm}_k}m_{\tilde{\chi}^{\pm}_l}
C_0(\tilde{u}_i,\tilde{\chi}^{\pm}_k,\tilde{\chi}^{\pm}_l) \right] \nonumber\\ 
&& 
-\frac{b^{C*}_{ik2}b^C_{il3}m_W^2}{e^2K^*_{ts}K_{tb} } 
U_{k1}U^*_{l1}
D_{27}(\tilde{u}_i,\tilde{\chi}^{\pm}_k,\tilde{\chi}^{\pm}_l,\tilde{l}^-) , 
\label{crnuch}
\\
C_{\nu,\tchi^0} &=& 
\frac{a^{N*}_{ik2}a^N_{jl3}}{2e^2K^*_{ts}K_{tb} } 
\left[ -\delta_{kl}(\Gamma_{DR})_{i\gamma}
(\Gamma_{DR}^{\dagger})_{\gamma j} C_{24}(\tilde{d}_i,\tilde{d}_j,
\tilde{\chi}^0_k) \right. \nonumber\\ 
&&  
+\delta_{ij}(N^*_{k3}N_{l3}-N^*_{k4}N_{l4})
\{ C_{24}(\tilde{d}_i,\tilde{\chi}^0_k,\tilde{\chi}^0_l) 
-\frac{1}{4} \}  \nonumber\\ 
&& \left.
+\frac{1}{2}\delta_{ij}(N_{k3}N^*_{l3}-N_{k4}N^*_{l4})
m_{\tilde{\chi}^0_k}m_{\tilde{\chi}^0_l}
C_0(\tilde{d}_i,\tilde{\chi}^0_k,\tilde{\chi}^0_l) \right] \nonumber\\ 
&& 
+\frac{a^{N*}_{ik2}a^N_{il3}m_W^2}{2e^2K^*_{ts}K_{tb} } 
\left[ \widetilde{N}^*_k\widetilde{N}_l
D_{27}(\tilde{d}_i,\tilde{\chi}^0_k,\tilde{\chi}^0_l,\tilde{\nu}) 
+\frac{1}{2}\widetilde{N}_k\widetilde{N}^*_l
m_{\tilde{\chi}^0_k}m_{\tilde{\chi}^0_l} 
D_0(\tilde{d}_i,\tilde{\chi}^0_k,\tilde{\chi}^0_l,\tilde{\nu}) 
\right] , \nonumber\\
\label{clnune}
\\
C'_{\nu,\tilde{\chi}^0} &=& 
\frac{b^{N*}_{ik2}b^N_{jl3}}{2e^2K^*_{ts}K_{tb} } 
\left[ \delta_{kl}(\Gamma_{DL})_{i\gamma}
(\Gamma_{DL}^{\dagger})_{\gamma j} C_{24}(\tilde{d}_i,\tilde{d}_j,
\tilde{\chi}^0_k) \right. \nonumber\\ 
&&  
-\delta_{ij}(N_{k3}N^*_{l3}-N_{k4}N^*_{l4})
\{ C_{24}(\tilde{d}_i,\tilde{\chi}^0_k,\tilde{\chi}^0_l) 
-\frac{1}{4} \}  \nonumber\\ 
&& \left.
-\frac{1}{2}\delta_{ij}(N^*_{k3}N_{l3}-N^*_{k4}N_{l4})
m_{\tilde{\chi}^0_k}m_{\tilde{\chi}^0_l}
C_0(\tilde{d}_i,\tilde{\chi}^0_k,\tilde{\chi}^0_l) \right] \nonumber\\ 
&& 
-\frac{b^{N*}_{ik2}b^N_{il3}m_W^2}{2e^2K^*_{ts}K_{tb} } 
\left[ \widetilde{N}_k\widetilde{N}^*_l
D_{27}(\tilde{d}_i,\tilde{\chi}^0_k,\tilde{\chi}^0_l,\tilde{\nu}) 
+\frac{1}{2} \widetilde{N}^*_k\widetilde{N}_l 
m_{\tilde{\chi}^0_k}m_{\tilde{\chi}^0_l} 
D_0(\tilde{d}_i,\tilde{\chi}^0_k,\tilde{\chi}^0_l,\tilde{\nu}) 
\right] , \nonumber\\
\label{crnune}
\\
C_{\nu,H^{\pm}} &=& \frac{h_t^2\cos^2\beta}{4e^2} 
\frac{x_{tH}}{(x_{tH}-1)^2}(1-x_{tH}+\log x_{tH} ), 
\label{clnuhp}
\\
C'_{\nu,H^{\pm}} &=& 
-\frac{(\hat{Y}_d)_{2\alpha}K^*_{t\alpha}K_{t\beta}(\hat{Y}_d)^*_{3\beta}
\sin^2\beta}{4e^2 K_{ts}^*K_{tb}} 
\frac{x_{tH}}{(x_{tH}-1)^2}(1-x_{tH}+\log x_{tH} ), 
\label{crnuhp}
\eea
where $x_{tH}=m_t^2/m_{H^{\pm}}^2$, $h_t=g_2m_t/(\sqrt{2}m_W\sin\beta)$. 
We assume flavor degeneracy in the lepton and slepton sectors. 
The formulas (\ref{clnugl}--\ref{crnuhp}) are derived 
from previous studies of the $b\to s\nu\bar{\nu}$ decays 
in the MSSM~\cite{BBMR,bsnunuSUSY}, as well as related works 
on the $K\to\pi\nu\bar{\nu}$ decays \cite{Kpinunu0,Kpinunu,Kpinunu2}. 
$C_{0,24}(a,b,c)\equiv C_{0,24}(m_a^2,m_b^2,m_c^2)$ and 
$D_{0,27}(a,b,c,d)\equiv D_{0,27}(m_a^2,m_b^2,m_c^2,m_d^2)$ are 
the three-point functions for the 
$Z$-penguin diagrams and four-point functions for the box diagrams, 
respectively \cite{PV}, in the convention of Ref.~\cite{PV2}. Ultraviolet 
divergence of $C_{24}$ cancels out in the formulas 
(\ref{clnugl}--\ref{crnune}). We ignore the masses of $(u,d,c)$ 
quarks, and include those of $(s,b)$ only when they are 
multiplied by $\tan\beta$. 
In this approximation, the neutral Higgs boson 
contributions to $C^{(')}_{\nu}$ vanish. 

The couplings and mixing matrices in Eqs.~(\ref{clnugl}--\ref{crnuhp}) 
are given as follows: 
The squark mixing matrices $(\Gamma_{QL},\Gamma_{QR})$ $(Q=U,D)$ give 
relations between the mass eigenstates 
$\sq_i=(\tilde{u}_i, \tilde{d}_i)(i=1-6)$ 
to the gauge eigenstates in the ``super-CKM'' basis 
$(\sq_{L\alpha},\sq_{R\alpha})(\alpha=1-3)$, 
which are related to the mass eigenbasis of the quarks 
$q_{\alpha}=[u_{\alpha}=(u,c,t), d_{\alpha}=(d,s,b)]$ by SUSY 
transformation, as 
\begin{equation}
\sq_{L\alpha} = (\Gamma_{QL}^{\dagger})_{\alpha j} \sq_j, 
\;\;\;\; 
\sq_{R\alpha} = (\Gamma_{QR}^{\dagger})_{\alpha j} \sq_j. 
\label{eq-gammasq}
\end{equation}
These matrices are determined to diagonalize the $6\times6$ 
mass matrices of squarks in the super-CKM basis, 
\bea 
M_{\sq}^2 &=&
\left( \begin{array}{cc}
M^2_{\tilde{Q} LL} & (M^2_{\tilde{Q} RL})^{\dagger} \\ 
M^2_{\tilde{Q} RL} & M^2_{\tilde{Q} RR}  
\end{array} \right) , \nonumber \\ 
(M^2_{\tilde{Q} LL})_{\alpha\beta} 
&=& (m^2_{\tilde{Q} LL})_{\alpha\beta} + (m_Q^{(0)})^{\dagger}(m_Q^{(0)})+
\delta_{\alpha\beta}(I_{3q_L}-e_q\sin^2\theta_W)m_Z^2\cos 2\beta, \nonumber \\
(M^2_{\tilde{Q} RR})_{\alpha\beta} 
&=& (m^2_{\tilde{Q} RR})_{\alpha\beta} + (m_Q^{(0)})(m_Q^{(0)})^{\dagger}+
\delta_{\alpha\beta}e_q\sin^2\theta_W m_Z^2\cos 2\beta, \nonumber \\
(M^2_{\tilde{U} RL})_{\alpha\beta} 
&=& (m^2_{\tilde{U} RL})_{\alpha\beta} - m_U^{(0)}\mu^* \cot\beta ,
\nonumber \\
(M^2_{\tilde{D} RL})_{\alpha\beta} 
&=& (m^2_{\tilde{D} RL})_{\alpha\beta} - m_D^{(0)} \mu^*\tan\beta .
\label{msq2}
\eea
In Eq.~(\ref{msq2}), off-diagonal elements of the 
soft SUSY breaking mass matrices $(m^2_{\tilde{Q} LL,RR,RL})$ induce 
flavor mixings which are not constrained by the CKM matrix in general, 
and may cause potentially large FCNC. 
$(m_Q^{(0)})_{\alpha\beta}$ are the ``bare'' mass 
matrices of the quarks. For the up-type squarks, it is just the running 
mass matrix $(m_U^{(0)})_{\alpha\beta}=(m_U)_{\alpha\beta} %
={\rm diag}(m_u,m_c,m_t)\sim {\rm diag}(0,0,m_t)$ 
in the standard model. For the down-type 
squarks, in contrast, $(m_D^{(0)})_{\alpha\beta}$ 
may substantially deviate from the standard model mass matrix 
$(m_D)_{\alpha\beta}={\rm diag}(m_d,m_s,m_b)$, as explained later. 
The quark-squark-chargino and quark-squark-neutralino couplings 
$(a^C_{ik\alpha},b^C_{ik\alpha},a^N_{ik\alpha},b^N_{ik\alpha})$ are then 
given in terms of the mixing matrices for squarks (\ref{eq-gammasq}),
for charginos $(V,U)$, and for neutralinos $N$ \cite{GH}, as 
\bea
a^C_{ik\alpha} &=& g_2 (\Gamma_{UL})_{i\beta} V_{k1}^* K_{\beta\alpha} 
- h_t (\Gamma_{UR})_{i3} V_{k2}^* K_{t\alpha} ,  \nonumber\\ 
b^C_{ik\alpha} &=&  -(\Gamma_{UL})_{i\beta} U_{k2} K_{\beta\gamma} 
(\hat{Y}_d)^*_{\alpha\gamma} ,    \nonumber\\ 
a^N_{ik\alpha} &=&  \sqrt{2}(-\frac{g_2}{2} N_{k2}^* +\frac{g_Y}{6}N_{k1}^*)
(\Gamma_{DL})_{i\alpha}  
+ (\hat{Y}_d)_{\beta\alpha} N_{k3}^* (\Gamma_{DR})_{i\beta},     \nonumber\\ 
b^N_{ik\alpha} &=&  \frac{\sqrt{2}g_Y}{3}N_{k1} (\Gamma_{DR})_{i\alpha}  
+ (\hat{Y}_d)^*_{\alpha\beta} N_{k3} (\Gamma_{DL})_{i\beta},  \label{eq-abino}
\eea 
Finally, $\widetilde{N}_k\equiv N_{k2}-\tan^2\theta_W N_{k1}$ in 
Eqs.~(\ref{clnune}, \ref{crnune}) denote the neutrino-sneutrino-neutralino 
couplings. 

We need some explanation for $(\hat{Y}_d)_{\alpha\beta}$, the bare Yukawa 
coupling matrix for down-type quarks. 
We start from the effective 
lagrangian for the couplings of $d_{iR}$ to the Higgs boson doublets 
($H_D$, $H_U$) in the MSSM, after integrating out the SUSY particles, 
\beq
{\cal L}_{\rm eff} = -(\hat{Y}_d)_{ij} \bar{d}_{iR} 
(d_{jL} H_D^0 - K^*_{kj}u_{kL} H_D^- )
-(\Delta Y_d)_{ij} \bar{d}_{iR} 
(d_{jL} H_U^{0*} + K^*_{kj}u_{kL} H_U^- )  + ({\rm h.c}). 
\label{eq-leff} 
\eeq
The couplings $(\Delta Y_d)_{ij}$ are forbidden at the tree-level by 
supersymmetry, but induced by SUSY particle loops with soft SUSY breaking. 
The running mass matrix in the standard model 
$(m_D)_{\alpha\beta}={\rm diag}(m_d,m_s,m_b)$ 
is then given by 
\bea
(m_D)_{\alpha\beta} &=& 
\frac{\sqrt{2}m_W}{g_2}\cos\beta[ \hat{Y}_d 
+\tan\beta \Delta Y_d  ] _{\alpha\beta} ,  \nonumber\\ 
&\equiv & [ m_D^{(0)} + \delta m_D  ] _{\alpha\beta} . 
\label{eq-md} 
\eea
Although the loop-generated $\Delta Y_d$ is suppressed 
relative to the tree-level coupling $\hat{Y}_d$, its 
contribution to $m_D$, $\delta m_D$, is enhanced by $\tan\beta$, 
as seen in Eq.~(\ref{eq-md}), and 
may become numerically comparable to the tree-level 
part $m_D^{(0)}\propto \hat{Y}_d$ at large $\tan\beta$ \cite{dmb}. 
On the other hand, the couplings of ($d_{iR}$, $\tilde{d}_{iR}$) to 
heavier Higgs bosons ($H^0$, $A^0$, $H^{\pm}$) and higgsinos 
$\tilde{H}_D$ are determined by $\hat{Y}_d$, as shown in 
Eqs.~(\ref{crnuhp}, \ref{eq-abino}), without $\tan\beta$-enhanced 
contributions from $\Delta Y_d$.
As a consequence, at large $\tan\beta$, 
these couplings may significantly deviate from the tree-level 
values \cite{carenaH0} given in terms of $(m_D)_{\alpha\beta}$ and, 
since $\Delta Y_d$ is not flavor diagonal in general, 
include flavor-mixing parts not determined by the CKM matrix, even 
in the super-CKM basis. 
The bare quark mass matrix $m_D^{(0)}$ should be also used 
in the mass matrix (\ref{msq2}) of the down-type squarks, 
which also receives no contributions from $\Delta Y_d$. 
The correction (\ref{eq-md}) therefore affects the masses and 
mixing matrices $(\Gamma_{DL},\Gamma_{DR})$ of the down-type squarks, 
generating additional flavor mixing for squarks. 
These $\tan\beta$-enhanced corrections to the down-type quarks and 
squarks are often comparable to the tree-level contributions 
in the MSSM at large $\tan\beta$, and should be included in realistic 
analysis of processes involving these particles \cite{carenaH0,BCRS}. 

Now we turn to the behavior of the SUSY and Higgs contributions 
(\ref{clnugl}--\ref{crnuhp}) to $(C_{\nu},C_{\nu}')$. The main 
part of these contributions comes from the $Z$ penguin diagrams 
through effective 
$Z_{\mu}\bar{s}_L\gamma^{\mu}b_L$ and $Z_{\mu}\bar{s}_R\gamma^{\mu}b_R$ 
vertices. 
Appearance of these vertices needs both the mixing 
between the second and third generations of quarks/squarks, and 
the SU(2)$\times$U(1) gauge symmetry breaking in the loops. 
For small or moderate value of $\tan\beta$, the largest SU(2) breaking 
in the loops are provided by the top quark and squarks. 
As a consequence, $C_{\nu,H^{\pm}}$ (\ref{clnuhp}) and 
$C_{\nu,\tchi^{\pm}}$ (\ref{clnuch}) are relevant. 
The former, however, is suppressed by 
$1/\tan^2\beta$ and only relevant for $\tan\beta\sim 1$, 
which is disfavored by experimental lower limit on the mass of 
the lightest Higgs boson. Therefore, only the latter, 
$C_{\nu,\tchi^{\pm}}$, is left 
as a potentially important SUSY contribution to $b\to s\nu\bar{\nu}$. 
Previous studies have shown \cite{bsnunuSUSY,bsnunusm4,bsnunusm3} 
that $C_{\nu,\tilde{\chi}^{\pm}}$ 
is enhanced by large $M_{\tilde{U}RL}^2$, 
especially by its flavor-mixing parts. Similar behavior is 
observed for the chargino contributions to the 
$K\to\pi\nu\bar{\nu}$ decays \cite{Kpinunu}. 

At large $\tan\beta$, however, other contributions to $b\to s\nu\bar{\nu}$ 
have the possibility to become sizable, by the following reasons: 
First, the SU(2)-breaking left-right mixing of the down-type squarks 
$(M_{\tilde{D}RL}^2)$ increases as $\tan\beta$ and may enhance 
the gluino contribution. 
Second, off-diagonal parts of the effective Yukawa coupling 
$\hat{Y}_d$ in Eq.~(\ref{eq-md}) induce the flavor-changing 
couplings of the down-type quarks, which are enhanced by $\tan\beta$ and 
not necessarily suppressed by the corresponding CKM matrix 
elements or quark masses. Especially, the element $(\hat{Y}_d)_{23}$, 
induced by flavor mixing in $M^2_{\tilde{D}RR}$, might give large 
Yukawa couplings of $s_R$ and enhance $C'_{\nu,H^{\pm}}$. This is similar 
to the case of $K\to\pi\nu\bar{\nu}$ at large $\tan\beta$ \cite{Kpinunu2}, 
where loop-induced couplings ($(\hat{Y}_d)_{13}$, $(\hat{Y}_d)_{23}$) give 
large effective $\bar{s}_Rd_RZ$ coupling. 
Therefore, the gluino (\ref{clnugl}, \ref{crnugl}) and 
charged Higgs boson (\ref{crnuhp}) contributions must be 
considered in the analysis of $b\to s\nu\bar{\nu}$ at large $\tan\beta$. 

\section{SUSY and Higgs contributions to $b\to s\nu\bar{\nu}$ and 
correlation with $b\to s\gamma$} 
\label{sect-MSSMcontr}

We present numerical results for the new physics contributions 
(\ref{clnugl}--\ref{crnuhp}) 
to the $b\to s\nu\bar{\nu}$ decay in the MSSM. We concentrate 
on the cases with large $\tan\beta$, which were not considered in 
previous studies. 

In the estimation of possible magnitudes of the new physics 
contributions (\ref{clnugl}--\ref{crnuhp}) to $b\to s\nu\bar{\nu}$, 
we need to take into account the 
constraints on SUSY and Higgs parameters from other FCNC processes. 
In this section, we consider the implication of the constraints 
from the radiative decay $b\to s\gamma$. 
This constraint is expected to be crucial since 
the SU(2)$\times$U(1) breaking and flavor mixing 
between quarks/squarks in the second and third generations, 
which are necessary to enhance the contributions to $b\to s\nu\bar{\nu}$, 
may also give large contributions to $b\to s\gamma$. 
Another reason to focus on $b\to s\gamma$ is the 
rather good agreement between experimental data \cite{bsgexp} and 
the standard model prediction \cite{bsgsm} of the inclusive branching ratio 
${\rm Br}(\bar{B}\to X_s\gamma)$. Indeed, the decay $b\to s\gamma$ in the 
MSSM have been shown \cite{BSGinSUSYproposal,BBMR,%
BSGlargeTBold,bsglargeTB2,LO-generalSUSY,NLO-SUSYa,NLO-SUSYb,BCRS} 
to give strong constraints 
on the Higgs and SUSY parameters. It should also be noted that 
the SUSY contributions to $b\to s\gamma$ are enhanced 
by $\tan\beta$ \cite{BSGlargeTBold,bsglargeTB2}. 

Here we do not attempt precise calculation of the experimental 
constraints from $b\to s\gamma$. 
Instead, we present a very rough estimation of 
the expected constraints in terms of the Wilson 
coefficients $(C_7,C'_7)(\mu)$ for $b\to s\gamma$, defined as 
\bea
&& H_{\rm eff} = 
 -\frac{4G_F}{\sqrt{2}}K^*_{ts}K_{tb}
 \left( C_7(\mu) {\cal O}_7(\mu) + C'_7(\mu) {\cal O}'_7(\mu) 
 \right) \,, \nonumber\\
&& {\cal O}_7 = \frac{e}{16\pi^2}m_b(\mu)
(\bar{s}_L\sigma^{\mu\nu}b_R)F_{\mu\nu}, \nonumber \\
&& {\cal O}'_7 = \frac{e}{16\pi^2}m_b(\mu)
(\bar{s}_R\sigma^{\mu\nu}b_L)F_{\mu\nu} .
\label{eq-bsg}
\eea

Below we show the correlations between 
$C^{(')}_{\nu}({\rm new})$, Eqs.~(\ref{clnugl}--\ref{crnuhp}), and 
new physics contributions to $C^{(')}_7$, $C^{(')}_7({\rm new})$, 
for each sector of new physics: namely, the gluino-squark, 
chargino-squark, and charged Higgs boson-top quark loop contributions, 
varying squarks mixing parameters which are relevant to $b\to s\nu\bar{\nu}$. 
For simplicity, we assume the flavor structures 
of the soft SUSY breaking terms in 
the squark mass matrices (\ref{msq2}) as 
\bea
m^2_{\tilde{Q} XX} &=& M_{\tilde{Q}}^2\left( 
\begin{array}{ccc} 1 & 0 & 0 \\ 0 & 1 & (\delta_{XX}^q)_{23} \\ 
0 & (\delta_{XX}^q)_{23} & 1 \end{array} \right) 
\;\; (XX=LL,RR) , \label{msq2LL} \\
m^2_{\tilde{U} RL} &=& m_t \left( 
\begin{array}{ccc} 0 & 0 & 0 \\ 0 & 0 & 0 \\ 
0 & (A_u)_{32} & (A_u)_{33} \end{array} \right) , \label{msu2RL}  
\eea 
Since CP violation is not essential for the analysis in this paper, 
all SUSY and Higgs parameters, including those in 
Eqs.~(\ref{msq2LL}, \ref{msu2RL}) are set to be real. 
We also set $m^2_{\tilde{D} RL}=0$ in Eq.~(\ref{msq2}) since 
its contribution to $M^2_{\tilde{D} RL}$ is, when 
the vacuum stability bounds \cite{stability} is applied, 
$O(m_b M_{\tilde{Q}})$ and subdominant compared to the second term 
$m_D^{(0)}\mu^*\tan\beta=O(m_b\tan\beta M_{\tilde{Q}})$. 
Note that the condition (\ref{msq2LL}) for $m^2_{\tilde{Q}LL}$
may be imposed only either 
$\tilde{Q}=\tilde{U}$ or $\tilde{Q}=\tilde{D}$, due to the 
SU(2) symmetry $(m_{\tilde{U}LL}^2)_{\alpha\beta}= %
K_{\alpha\gamma}(m_{\tilde{D}LL}^2)_{\gamma\delta}K^*_{\beta\delta}$. 

We calculate the new physics contributions to $C_{\nu}^{(')}$ and 
$C^{(')}_7$ at the leading one-loop order (see 
Refs.~\cite{BBMR,BSGlargeTBold,bsglargeTB2,LO-generalSUSY} for the 
formulas of $C^{(')}_7$), but improved by including the $\tan\beta$-enhanced 
corrections to the quark/squark Yukawa couplings from Eq.~(\ref{eq-md}) and, 
for $C^{(')}_7$, also from the proper vertex corrections\footnote{These 
vertex corrections also appear in $C_{\nu,H^{\pm}}$. However, we ignored 
the corrections in Eq.~(\ref{clnuhp}), since $C_{\nu,H^{\pm}}$ itself is 
strongly suppressed by $1/\tan^2\beta$ and numerically negligible.} 
to the $u_{iR}$ couplings to $(H^0,A^0,H^{\pm})$ \cite{NLO-SUSYa,NLO-SUSYb}, 
in the effective lagrangian formalism \cite{NLO-SUSYa}. 
In these formulas, we use the running quark masses and $\alpha_s$ 
at the renormalization scale $\mu=M_{\tilde{Q}}$, 
calculated from $m_t({\rm pole})=171$ GeV, $m_b(m_b)=4.2$ GeV, 
$m_s(2 {\rm GeV})=95$ MeV, $m_q({\rm others})=0$ and $\alpha_s(m_Z)=0.12$, 
which give $C^{(')}_{\nu}(\mu)$ and $C^{(')}_7(\mu)$ 
at the renormalization scale $\mu=M_{\tilde{Q}}$. 
For SUSY and Higgs parameters, we fix the following parameters: 
$\tan\beta=50$, $M_{\tilde{Q}}=500$ GeV, $m_{\sg}=500$ GeV, 
$M_2=300$ GeV, $M_1=150$ GeV, while varying other parameters. 
We also impose the bounds $m_{\tchi^{\pm}}>100$ GeV and 
$m_{\sq}>250$ GeV, suggested by experimental search limits for 
SUSY particles. 

For each sector of the new physics, rough estimates of the 
bounds on the contributions to $(C_{\nu},C'_{\nu})$ are obtained 
by requiring that the magnitudes of $(C_7,C'_7)({\rm new})$ 
should be smaller than the standard model contribution 
$C_{7,\rm SM}(\mu\sim m_W)\sim -0.2$. 

\subsection{Gluino contributions}
\label{subsect-gluino}

\begin{figure}[th] 
\begin{center} 
\includegraphics[width= 14cm]{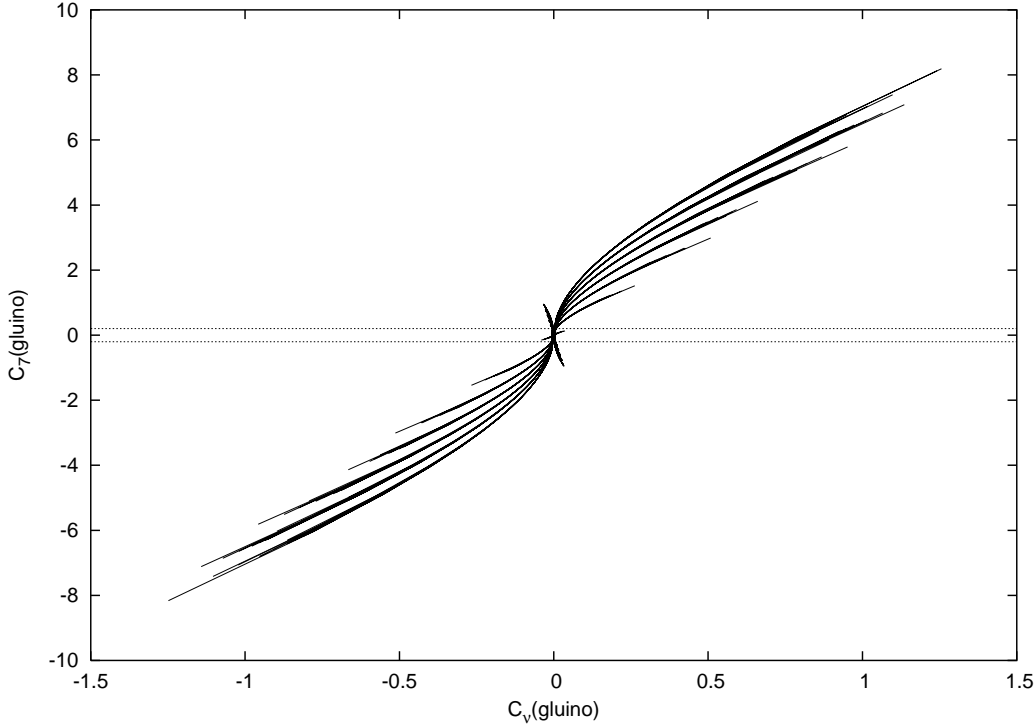} 
\end{center} 
\vspace{-0.1truecm}
\caption[f1]{\small 
Correlation between $C_{\nu,\sg}$ and $C_{7,\sg}$. Parameters are 
$\tan\beta=50$, $\mu=[-550, 550]$ GeV, 
$(\delta_{LL,RR}^d)_{23}=[-0.3, 0.3]$. 
Other parameters are given in the text. 
Horizontal lines indicate the region $|C_{7,\sg}|<0.2$. 
} 
\label{fig-gluino}
\end{figure}

The gluino-squark contributions $C^{(')}_{\nu,\sg}$ are 
induced by the flavor and left-right mixing of the down-type squarks. 
In Fig.~\ref{fig-gluino}, the gluino contribution $C_{\nu,\sg}$ 
is shown as a correlation with $C_{7,\sg}$, for 
parameter scan over 
$(\delta^d_{LL})_{23}=[-0.3, 0.3]$, $(\delta^d_{RR})_{23}=[-0.3,0.3]$, and 
$\mu=[-550,550]$ GeV. $(A_u)_{33}$ and $(A_u)_{32}$ are set to 0. 
Correlation between $C'_{\nu,\sg}$ and $C'_{7,sg}$ for the same parameters 
is obtained from Fig.~\ref{fig-gluino} by changing the sign of 
the horizontal axis. 
Large $|C_{\nu,\sg}|$ is obtained for large negative $\mu$ 
and large $(\delta_{LL,RR})_{23}$, which cause large 
$\tilde{b}_R-\tilde{s}_L$ mixing. 
It is seen that $|C_{\nu,\sg}|$ can be larger than $1$, which 
gives about $30$~\% correction to the 
standard model prediction of the decay width (\ref{eq-inclwidth}).
However, by requiring 
$|C_{7,\tilde{g}}|<|C_{7,{\rm SM}}(\mu_W)|\sim 0.2$, 
magnitudes of $C_{\nu,\tilde{g}}$ are constrained to be 
much smaller than $C_{\nu,\rm SM}\sim -6.8$. 
Therefore, without very precise cancellation between new physics contributions 
to $b\to s\gamma$, gluino contributions to $b\to s\nu\bar{\nu}$ should be 
completely negligible, even for $\tan\beta\gg 1$, to satisfy the bound 
from $b\to s\gamma$. 

Here we briefly comment on the neutralino contributions 
$C^{(')}_{\nu,\tchi^0}$, Eqs.~(\ref{clnune}, \ref{crnune}). 
Similar to the gluino contributions, 
$C^{(')}_{\nu,\tchi^0}$ are induced by the $\tilde{b}-\tilde{s}$ mixing 
in the loops. However, due to small couplings, these contributions 
are much smaller than the gluino contributions $C^{(')}_{\nu,\sg}$ 
for most parameter regions and therefore not discussed here. 

\subsection{Chargino contributions}
\label{subsect-chargino}

The chargino-squark loop contributions $C_{\nu,\tilde{\chi}^{\pm}}$, 
Eq.~(\ref{clnuch}), have been studied 
in previous works \cite{bsnunuSUSY,bsnunusm4} at small or moderate 
value of $\tan\beta$. In these works, it has been shown 
that they might give sizable contributions, larger than 
the uncertainty of the standard model predictions (\ref{eq-bsnunusm}), 
for large flavor-mixing element of $m_{\tilde{U}RL}^2$ in Eq.~(\ref{msu2RL}), 
especially its ($\tilde{t}_R$, $\tilde{c}_L$)-mixing 
element $(m_{\tilde{U}RL}^2)_{32}\sim (A_u)_{32}m_t$. 

\begin{figure}[th] 
\begin{center} 
\includegraphics[width= 14cm]{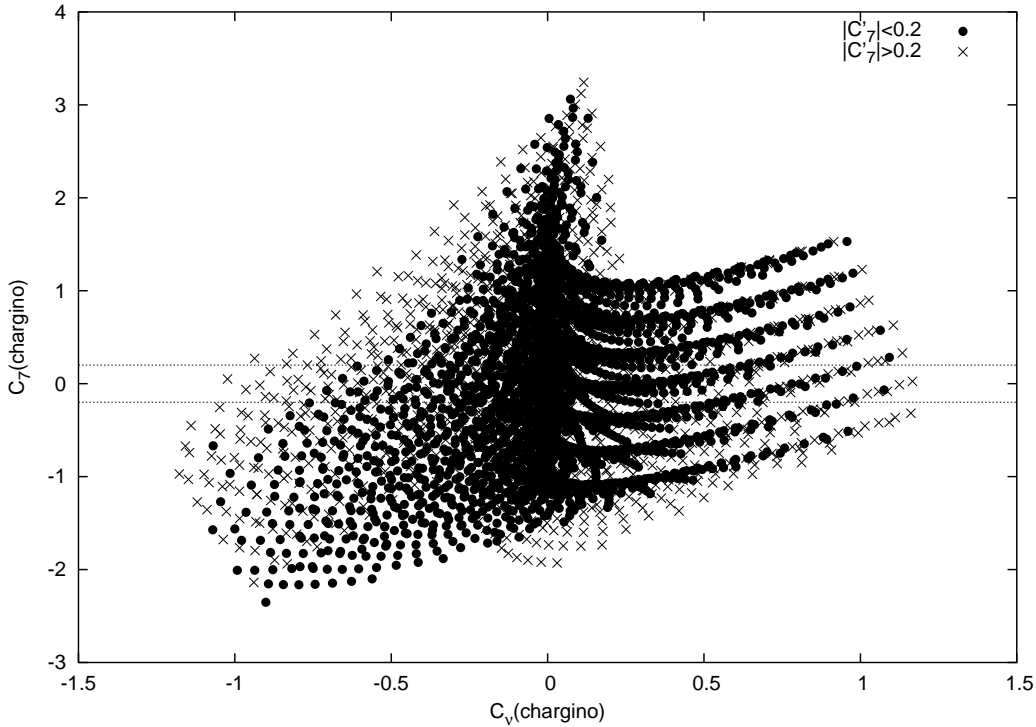} 
\end{center} 
\vspace{-0.1truecm}
\caption[f1]{\small 
Correlation between $C_{\nu,\tilde{\chi}^{\pm}}$ and 
$C_{7,\tilde{\chi}^{\pm}}$ for parameters 
$(A_u)_{33}=[-1500,1500]$ GeV, $(A_u)_{32}=[-1500,1500]$ GeV, 
and $(\delta_{LL}^u)_{23}=[-0.3, 0.3]$. The points with 
$|C'_{7,\tilde{\chi}^{\pm}}|$ smaller (larger) than 
$|C_{7,\rm SM}|\sim 0.2$ are denoted by dots (crosses). 
Other parameters are set as in the text. 
} 
\label{fig-chargino}
\end{figure}

Figure \ref{fig-chargino} shows the correlation between 
$C_{\nu,\tchi^{\pm}}$ and $C^{(')}_{7,\tchi^{\pm}}$, for varying 
parameters over 
$(A_u)_{33}=[-1500,1500]$ GeV, $(A_u)_{32}=[-1500,1500]$ GeV, 
and $(\delta_{LL}^u)_{23}=[-0.3, 0.3]$. 
Other parameters are fixed at 
$\mu=500$ GeV, $m_{\tilde{l}_L^{\pm}}=400$ GeV, 
and $(\delta_{RR}^u)_{23}=0$. For these parameters, 
$C'_{\nu,\tilde{\chi}^{\pm}}$ is negligibly small ($<0.02$) 
and not shown here. 
As is the case of the gluino contributions, SUSY parameters 
which give large $C_{\nu,\tilde{\chi}^{\pm}}$ tend to also give 
large $C^{(')}_{7,\tilde{\chi}^{\pm}}$. 
The resulting constraint on $C_{\nu,\tilde{\chi}^{\pm}}$ gets tighter 
as $\tan\beta$ increases, 
since $C^{(')}_{7,\tilde{\chi}^{\pm}}$ are 
enhanced by $\tan\beta$ while $C_{\nu,\tilde{\chi}^{\pm}}$ is not. 
Nevertheless, the correlation is not so strong as in the gluino sector, 
as seen in Fig.~\ref{fig-chargino}. 
This is due to 
the different dependences of $C_{\nu,\tilde{\chi}^{\pm}}$ 
and $C^{(')}_{7,\tilde{\chi}^{\pm}}$ on two A-term elements, 
$(A_u)_{33}$ and $(A_u)_{32}$ in Eq.~(\ref{msu2RL}). 
In fact, as seen in Fig.~\ref{fig-chargino}, we may 
have $|C_{\nu,\tilde{\chi}^{\pm}}|>1$ while keeping 
$|C^{(')}_{7,\tilde{\chi}^{\pm}}|<0.2$. Even larger value of 
$C_{\nu,\tilde{\chi}^{\pm}}$ might be possible by 
careful choice of the SUSY parameters. 
The resulting deviations of the decay widths from the 
standard model predictions (\ref{eq-bsnunusm}) could 
be proved at future B factories, if the theoretical uncertainties 
of the exclusive widths in Eq.~(\ref{eq-bsnunusm}), mainly coming from 
the meson form factors, are reduced. 
However, one must note that the large chargino contribution 
is realized by the fine tuning between SUSY parameters, especially 
$(A_u)_{33}$ and $(A_u)_{32}$, to realize 
small $C^{(')}_{7,\tilde{\chi}^{\pm}}$. 

\subsection{Charged Higgs boson contributions}
\label{subsect-Higgs}

\begin{figure}[th] 
\begin{center} 
\includegraphics[width= 14cm]{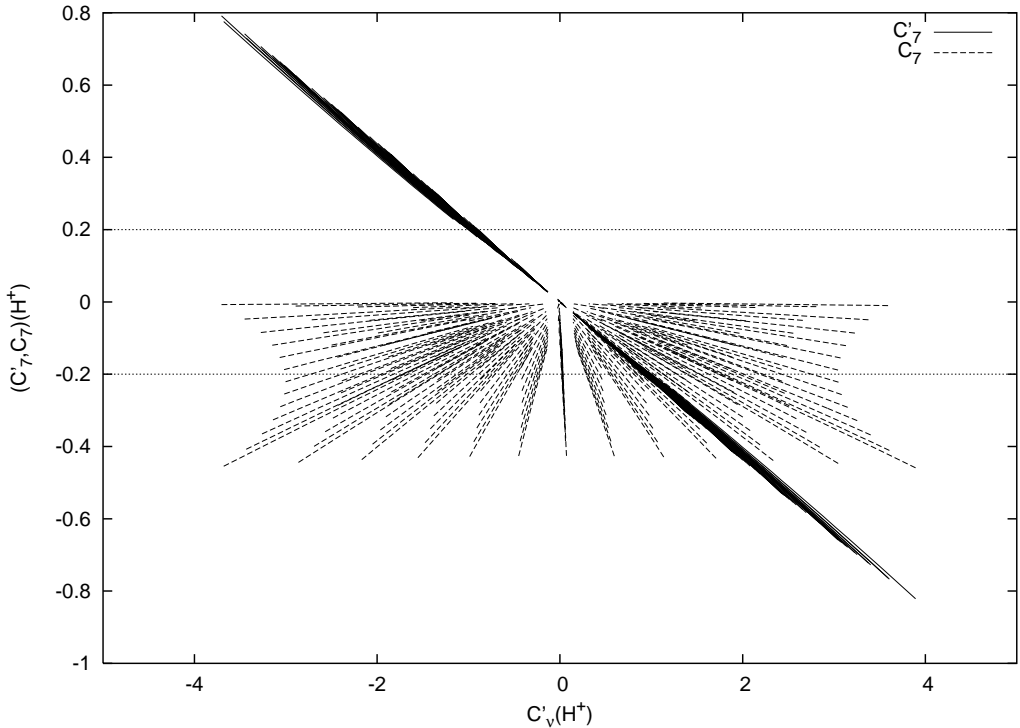} 
\end{center} 
\vspace{-0.1truecm}
\caption[f1]{\small 
Correlation between $C_{\nu,H^{\pm}}$ and 
$C^{(')}_{7,H^{\pm}}$ at $\tan\beta=50$, 
$(\delta_{LL,RR}^d)_{23}=[-0.3,0.3]$, and $m_{H^{\pm}}=[400,1000]\,$GeV. 
Other parameters are given in the text. 
} 
\label{fig-Hp}
\end{figure}

As discussed in the previous section, only $C'_{\nu,H^{\pm}}$, 
Eq.~(\ref{crnuhp}), 
is relevant at large $\tan\beta$. This contribution comes from 
the $H^-\bar{s}_Rt_L$ coupling 
$\sim (\hat{Y}_d)_{2\alpha}K^*_{t\alpha}\sim (\hat{Y}_d)_{23}$, which is 
generated by the flavor mixing involving $\tilde{s}_R$ 
through the $\tan\beta$-enhanced loop corrections (\ref{eq-md}). 
In Fig.~\ref{fig-Hp}, we show the correlations between 
$C'_{\nu,H^{\pm}}$ and $C^{(')}_{7,H^{\pm}}$ at 
$\mu=-500$ GeV, $(A_u)_{33}=0$ GeV, $(A_u)_{32}=0$ GeV, 
$(\delta_{LL}^d)_{23}=[-0.3,0.3]$, $(\delta_{RR}^d)_{23}=[-0.3,0.3]$, 
and $m_{H^{\pm}}=[400,1000]$ GeV. 
In contrast to the gluino and chargino contributions to $b\to s\gamma$, 
the main parts of $C^{(')}_{7,H^{\pm}}$ are not enhanced 
by $\tan\beta$. Moreover, 
the correlations between $C'_{\nu,H^{\pm}}$ and 
$C_{7,H^{\pm}}$ is severely affected by different 
parameter dependences of two generation-mixing $H^{\pm}$ couplings: 
the effective $\bar{s}_Rt_LH^-$ coupling $\sim(\hat{Y}_d)_{23}$ 
in $C'_{\nu,H^{\pm}}$ and $C'_{7,H^{\pm}}$, and 
$O(\tan\beta)$ proper vertex corrections to the 
$\bar{s}_Lt_RH^-$ coupling \cite{NLO-SUSYa,NLO-SUSYb} in $C_{7,H^{\pm}}$. 
As a result, similar to the case of chargino contributions, 
there is possiblity to have sizable $C'_{\nu,H^{\pm}}$ while 
keeping $C^{(')}_{7,H^{\pm}}$ small. 

\section{Constraint from $B_s\to\mu^+\mu^-$ on the $H^{\pm}$ contribution} 
\label{sect-bsmumu}

In addition to $b\to s\gamma$, several other $b\to s$ FCNC processes 
have been measured in recent experiments. 
Since most of these measurements show rather good consistency with 
the standard model predictions, they should give 
additional constraints on the SUSY and Higgs parameters, 
and their contributions to $b\to s\nu\bar{\nu}$. 
For example, measurements of the 
$B_s-\bar{B}_s$ oscillation \cite{bsbsbarexp} impose constraints 
on the $\tilde{b}-\tilde{s}$ mixing, 
especially on $(\delta^d_{LL})_{23}$ 
and $(\delta^d_{RR})_{23}$ \cite{bsbsbarSUSY}. 
Here we just show a case of these constraints: implication 
of the upper limit of the branching ratio for $B_s\to\mu^+\mu^-$ 
on the $H^{\pm}$ contributions $C'_{\nu,H^{\pm}}$, at large $\tan\beta$. 

As seen in Eq.~(\ref{crnuhp}), large value of $C'_{\nu,H^{\pm}}$ is 
obtained for parameters which give large 
effective $H^- \bar{s}_R t_L$ Yukawa coupling $\sim (\hat{Y}_d)_{23}$. 
As discussed in Sect.~\ref{sect-bsnunu}, the parameter 
$(\hat{Y}_d)_{23}$ also gives 
the flavor-changing $(H^0,A^0)\bar{s}_Rb_L$ couplings 
of the heavier neutral Higgs bosons $(H^0,A^0)$. 
On the other hand, at large $\tan\beta$, 
this coupling gives ``tree-level'' contributions to 
the $B_s\to\mu^+\mu^-$ decay by the Higgs penguin 
diagrams \cite{btomumuDP,BCRS}, 
which are often much larger than the standard model contributions 
by orders of magnitude. Requiring that these Higgs penguin contributions 
do not saturate the experimental upper bound 
${\rm Br}(B_s\to \mu^+\mu^-)<10^{-7}$ at 95\% C.L. \cite{bstomumuexp}, 
and neglecting mass difference between ($H^0$, $A^0$), 
the condition 
\beq
|(\hat{Y}_d)_{32}|^2+|(\hat{Y}_d)_{23}|^2 <0.2\cos^2\beta 
(m_A/500\,{\rm GeV})^4, 
\label{eq-bsmumubound} 
\eeq
is imposed on the $b-s$ mixing Yukawa couplings 
($(\hat{Y}_d)_{32}$, $(\hat{Y}_d)_{23}$) at the renormalization scale 
$\mu_b\sim m_b$. In the approximation of neglecting the QCD running 
between $\mu_b$ and $M_{\tilde{Q}}$, and also the $O(\tan\beta)$-enhanced 
correction to the $\bar{t}_Lb_RH^+$ coupling $\sim(\hat{Y}_d)_{33}$, 
Eq.~(\ref{eq-bsmumubound}) implies the bound 
$|C'_{\nu,H^{\pm}}|<0.15$ for $\tan\beta=50$ and $m_A<1000$ GeV, 
which is completely negligible compared to $C_{\nu,\rm SM}$. 
We expect that this strong constraint still holds when more 
rigorous estimation of $B_s\to\mu^+\mu^-$ is adopted. 

\section{Conclusion}
\label{sect-conclu}

We have studied the flavor-changing decay $b\to s\nu\bar{\nu}$ 
in the MSSM, at large $\tan\beta$ and with general flavor mixing of 
squarks. This case is interesting since the 
gluino and $H^{\pm}$ loops, which are negligible at moderate value of 
$\tan\beta$ and with minimal flavor violation for squarks, 
are enhanced and might give contributions to this decay, 
comparable to the standard model and chargino loop contributions. 
This is due to the $\tan\beta$-enhanced SU(2)$\times$U(1) gauge 
symmetry breaking and flavor mixing in the down-type squark sector, 
and loop-generated effective flavor-changing couplings 
of the charged Higgs boson to quarks and squarks. 
However, the contributions to $b\to s\nu\bar{\nu}$ by new physics 
should be constrained by experimental data for other $b\to s$ processes. 

In this paper, we have focused our attention to the constraints from 
the radiative decay $b\to s\gamma$, since both of the 
$b\to s\nu\bar{\nu}$ and $b\to s\gamma$ decays are enhanced by 
the SU(2)$\times$U(1) symmetry breaking and 
flavor mixing between the second and 
third generations of the quarks/squarks in the loops. 
As a very rough estimation of the constraints by $b\to s\gamma$, 
we have calculated the correlations between new physics 
contributions to the Wilson coefficients $C^{(')}_{\nu}$ and 
$C^{(')}_7$ for the $b\to s\nu\bar{\nu}$ and $b\to s\gamma$ decays, 
respectively, for each new physics sector: 
gluino-squark, chargino-squark, and charged Higgs-quark 
loops. Calculation has been done at the leading order, but 
including $\tan\beta$-enhanced corrections to the quark Yukawa 
couplings in the loops. 
It has been demonstrated that the requirement that the new physics 
contributions $C^{(')}_7({\rm new})$ for each sector 
are smaller than $C_{7,\rm SM}$ strongly constrains 
the new physics contributions $C^{(')}_{\nu}({\rm new})$. Especially, 
the gluino contributions $C^{(')}_{\nu,\sg}$ are suppressed 
much below $C_{\nu,\rm SM}$ due to their strong correlation with 
$C^{(')}_{7,\sg}$. 
In contrast, although the constraints by $C^{(')}_7$ are 
also tight for chargino and charged Higgs boson contributions, 
there still remains a possibility that their contributions to 
$C^{(')}_{\nu}$ become sizable, $O(10)$\%, while keeping contributions 
to $C^{(')}_7$ below $C_{7,\rm SM}$. 

As an example of the constraints by other $b\to s$ processes, we 
have also considered the Higgs penguin contributions to 
the decay $B_s\to\mu^+\mu^-$, which might become much larger than 
the standard model contribution at large $\tan\beta$. It has been shown 
that the present experimental upper bound of the decay ratio may 
impose strong constraints on $C'_{\nu,H^{\pm}}$, suppressing it 
much below $C_{\nu,\rm SM}$. 

For more realistic analysis of $b\to s\nu\bar{\nu}$ in the MSSM and 
estimation of the new physics contributions, we need to scan 
over wider parameter space, including correlations between 
different contributions to $C^{(')}_{\nu}$, 
main parts of the QCD corrections and hadronic effects, and 
constraints from other flavor-changing processes using more precise 
formulas of the new physics contributions. We leave such studies 
for future works. 

\vspace*{0.5truecm}
\noindent 
{\bf Acknowledgements}  

The author thanks Francesca Borzumati for earlier collaboration. 
The work was supported in part by the Grant-in-Aid for 
Scientific Research on Priority Areas 
from the Ministry of Education, Culture, Sports, 
Science and Technology of Japan, No.~16081202 and 
17340062. 

%

\end{document}